\documentclass[a4paper,10pt,twocolumn,final,conference,oneside]{IEEEtran}%
\IEEEoverridecommandlockouts

\usepackage{tikz} 
\usepackage[american]{circuitikz} 
\usetikzlibrary{shapes.geometric, arrows, matrix, calc, shapes}

\usepackage[free-standing-units]{siunitx}
\usepackage{steinmetz}
\usepackage{multirow, bigdelim}
\graphicspath{{./}{imgs/}}

\usepackage{amsmath}
\usepackage{amssymb}
\usepackage{stackrel}

\usepackage{icomma}

\usepackage[utf8]{inputenc}
\usepackage[T1]{fontenc}

\usepackage{graphicx} 	          

\usepackage{float}		

\usepackage{color}
\usepackage{threeparttable}
\usepackage{mathtools}

\usepackage{cite}

\makeatletter
\ctikzset{lx/.code args={#1 and #2}{ 
  \pgfkeys{/tikz/circuitikz/bipole/label/name=\parbox{1cm}{\centering #1  \\ #2}}
    \ctikzsetvalof{bipole/label/unit}{}
    \ifpgf@circ@siunitx 
        \pgf@circ@handleSI{#2}
        \ifpgf@circ@siunitx@res 
            \edef\pgf@temp{\pgf@circ@handleSI@val}
            \pgfkeyslet{/tikz/circuitikz/bipole/label/name}{\pgf@temp}
            \edef\pgf@temp{\pgf@circ@handleSI@unit}
            \pgfkeyslet{/tikz/circuitikz/bipole/label/unit}{\pgf@temp}
        \else
        \fi
    \else
    \fi
}}

\ctikzset{lx^/.style args={#1 and #2}{ 
    lx=#2 and #1,
    \circuitikzbasekey/bipole/label/position=90 } 
}

\ctikzset{lx_/.style args={#1 and #2}{ 
    lx=#1 and #2,
    \circuitikzbasekey/bipole/label/position=-90 } 
}
\makeatother

\ifCLASSINFOpdf
  
\else
  
\fi

\usepackage{epstopdf}

\usepackage{xcolor}
\usepackage{booktabs}
\usepackage{multirow}
\usepackage{longtable}
\usepackage{array}
\usepackage{tabularx}
\usepackage{makecell}
\usepackage{hyperref}
\usepackage{pict2e}

\makeatletter
\newcommand\doslashcirc[2]{%
	\sbox\z@{$#1\m@th\circ$}%
	\setlength\unitlength{\wd\z@}
	\begin{picture}(1,1)
	\roundcap
	\put(0,0){\box\z@}
	\put(0,0){\line(1,1){1}}
	\end{picture}%
}
\makeatother
\usepackage{wasysym}

\hypersetup{
	colorlinks=true,
	linkcolor=black,
	filecolor=magenta,
	urlcolor=cyan,
	citecolor=black
}

\newcolumntype{L}[1]{>{\raggedright\arraybackslash}p{#1}}
\newcolumntype{C}[1]{>{\centering\arraybackslash}p{#1}}
\newcolumntype{Y}[1]{>{\raggedright\arraybackslash}p{#1}}

\newcommand{\drawpulse}[2][0]{
	\begin{scope}[rotate=#1]
		\draw (#2.center) ++(-3mm, -2mm) -| ++(2mm,5mm)
		-- ++(2mm,0mm) |- ++(2mm, -5mm);
	\end{scope}
}

\begin{document}
%
\ctikzset{
	american resistors,
	american inductors,
	full diodes,
	bipoles/cuteswitch/thickness=0.3,
	diodes/scale=0.5,
	resistors/scale=0.7,
	capacitors/scale=0.7,
	capacitors/thickness=4,}

\title{Wireless Power Transfer in Titanium Implants}


\makeatletter
\newcommand{\linebreakand}{%
\end{@IEEEauthorhalign}
\hfill\mbox{}\par
\mbox{}\hfill\begin{@IEEEauthorhalign}
}
\makeatother

\author{\IEEEauthorblockN{R. W. Porto$^1$, L. Murliky$^2$, F. R. de Sousa$^3$,~\IEEEmembership{Senior Member,~IEEE}, A. S. de Almeida$^4$,\\ H. M. de Albuquerque$^4$, V. J. Brusamarello$^2$,~\IEEEmembership{Senior Member,~IEEE}}
	\IEEEauthorblockA{
		\textit{$^1$Instituto Federal do Rio Grande do Sul} IFRS, Porto Alegre, Brasil\\ 
		\textit{$^2$Universidade Federal do Rio Grande do Sul} UFRGS, Porto Alegre, Brasil - PPGEE \\
		\textit{$^3$Universidade Federal de Santa Catarina} 
		UFSC, Florian\'{o}polis, Brasil - PPGEEL\\
		\textit{$^4$MSC MED Engenharia e Tecnologia M\'{e}dica Ltda}\\ 
		rodrigo.porto@restinga.ifrs.edu.br, rangel@ieee.org, valner.brusamarello@ufrgs.br}
}


%


\maketitle

\begin{abstract}
.
This work addresses the problem of applications that require wireless power transfer to devices embedded within conductive blocks. The emergence of eddy currents tends to generate a magnetic field opposing that of the transmitter, drastically reducing the coupling coefficient and, consequently, the efficiency of the process. A case study involving a titanium implant is presented and analyzed with a load designed to ensure a constant current of 1 mA. An experimental setup is assembled and evaluated to compare differences in the measured electrical parameters when the receiver is embedded in the implant, considering several solutions that mitigate or avoid the effects of eddy currents. The results show that such a system may be technically unfeasible if careful attention is not paid to the geometry of the cavity in which the receiver is positioned.

\end{abstract}



%
\IEEEpeerreviewmaketitle

\section{Introduction}

Wireless power transfer (WPT) can be achieved through different methods and physical mechanisms. In this context, electromagnetic techniques stand out, including low-frequency methods employing inductive or capacitive coupling, high-frequency approaches using RF, as well as optical methods, among others. In the frequency range from a few kilohertz to several megahertz, inductive coupling has been the most widely adopted approach in applications such as battery charging for electric vehicles and medium-power systems \cite{9248590,brusamarello}. It has also been extensively used for charging batteries in remote Instrumentation electronic systems \cite{10412337,8410809,8038802,8006394}, wearables \cite{8968384} and in human body implants, such as pacemakers \cite{10979533} and brain implants \cite{7328297}.

Such energy transfer mechanisms rely on near-field electromagnetic coupling and are highly dependent on the properties of the materials surrounding the application. Specifically, non-conductive or weakly conductive materials--such as plastics, wood, and even biological tissues--tend to have negligible effects. However, when conductive materials are introduced, induced currents tend to cancel the electromagnetic flux, thereby reducing or even nullifying the energy transfer process \cite{7325300}.

Near field electromagnetics are characterized, among other properties, by the absence of propagation or radiation. Nevertheless, energy can remain confined within a coupled circuit. In the presence of a conductor, eddy currents arise, as defined by Faraday's law, whose direction (according to Lenz's law) opposes the variation of the magnetic flux that generated them. Therefore, when a magnetic field impinges upon a conductor, not only energy is lost through dissipation, but the resulting incident field is also attenuated. Thus, wireless power transmission via magnetic coupling can become highly inefficient under certain conditions where conductors are present in the vicinity.

In this context, medical implants are commonly made from conductive materials such as titanium \cite{implante1}, \cite{implante2}. In scenarios where electromagnetic fields are present, or in cases where wireless energy delivery to nearby electronic devices is desired, eddy currents pose a significant challenge. These currents can severely degrade the magnetic coupling efficiency, thereby reducing the overall effectiveness of the power transmission process \cite{heat_implant}.

This article addresses the challenge of delivering power via inductive coupling to an electronic device located in a cavity of a titanium alloy implant -- or a similar conductive material. The implant is designed according to strict dimensional and geometric constraints, which must be preserved. As a result, the receiving coil and embedded electronics must conform to these limitations. This study demonstrates that, depending on the power required by the electronic device for measurement and data transmission tasks, as well as on the imposed mechanical and geometric constraints, the system design may become unfeasible. Eddy currents induced in the conductive structure tend to cancel the magnetic flux passing through the receiving coil, thereby reducing its mutual inductance or magnetic coupling. The problem is initially analyzed and simulated, followed by the presentation of different solutions based on coil geometry modifications and the use of ferromagnetic materials in the core and/or surrounding the device. Finally, different coil geometries and structural configurations that enhance magnetic coupling are proposed, along with selected case studies supported by experimental results.

\section{Eddy Currents on Implants} 
\label{modelo}

Titanium (Ti) is a metallic element with proven biocompatibility. Ti also exhibits high corrosion resistance, a high strength-to-weight ratio, and a low elastic modulus. These properties make Ti one of the most used materials in the fabrication of components and mechanical structures implanted in the human body \cite{implante1}, as well as in instruments and medical equipment employed in surgical procedures \cite{10429831}, and in coatings for devices such as cardiac pacemakers \cite{1548400}.

Ti has an electrical conductivity of 3.1\% IACS (International Annealed Copper Standard), corresponding to $1.798\times10^6$ S/m. In the presence of electromagnetic fields, eddy currents are induced in these structures. Both in the case of prostheses and pacemakers, such currents may cause heating of the body. In wireless power transmission systems, they lead to a reduction in process efficiency.

According to Faraday's law, a time-varying magnetic field induces an electric field $\textbf{E}$ \cite{168666}.

\begin{equation}
\mathcal{E}=-\frac{d\Phi}{dt} = \oint_C \mathbf{E} \cdot \mathrm{d}\mathbf{l}
\label{eq1}
\end{equation}
where $\mathcal{E}$ is the electromotive force and $\Phi$ is the magnetic flux through the circuit or conductor. Therefore, in each conductor, a current density arises:
\begin{equation}
J_{eddy} = \sigma E
\label{densidade}
\end{equation}
(where $\sigma$ is the electrical conductivity of the material); this term expresses the presence of parasitic (eddy) currents in conductive media, which generate heat $Q$ (Joule's law):
\begin{equation}
Q = \sigma^{-1} J.J
\end{equation}
and the Lorentz forces: 
\begin{equation}
f_l = J \times B    
\end{equation}

In the context of WPT, the Joule effect causes device heating, and if the receiving coil is located near or even embedded within the conductor, the parasitic (eddy) currents generate a magnetic field opposite to the one that originated them (Lenz's law). This reduces the magnetic flux and, consequently, the magnetic coupling and the energy received by the receiver. Therefore, in such cases, it is desirable to minimize eddy currents and their effects.

One way to reduce induced currents is to divide the solid core into partitions. In \cite{7123650}, a core structure composed of rods (each surrounded by an insulating layer and aligned parallel to the magnetic field direction) is compared with both a solid core and a laminated core with stacked partitions. The rod-type core is a structure with $n \times m$ partitions. It was found that when the laminated core and the multi-rod core contain the same amount of iron and insulating material, the multi-rod configuration produces fewer eddy currents. Moreover, the reduction of eddy currents in the multi-rod core leads to lower magnetic circuit reluctance, which in turn results in reduced copper losses due to the corresponding decrease in magnetizing current.

\subsection{Linear Equivalent Circuit Model}

The presence of conductors being penetrated by electromagnetic fields satisfies (\ref{eq1}), provided that the material does not saturate or operates within a linear region. The magnetic flux $\Phi$ defined in (\ref{eq1}) can be expressed as $\Phi = \int B \cdot dA$, where $B$ is the magnetic flux density, which in turn depends on the magnetization $M$:

\begin{equation}
B=\mu_o(H+M)
\end{equation} where $H$ is the applied magnetic field intensity. The $B \times H$ hysteresis curve of materials illustrates these relationships and the regions of nonlinearity. Fig.~\ref{cir_eq1} shows an equivalent linear electrical circuit model composed of the transmitting coil with impedance $Z_T$, the receiving coil with impedance $Z_R$ connected to a load $Z_L$, and a coil representing the induced currents with impedance $Z_e=R_e+j\omega L_e$. The magnetic couplings between the coils are represented by three mutual inductances: between the transmitter and receiver, $M_{TR}$; between the transmitter and the material (generating eddy currents), $M_{Te}$; and between the receiver circuit and the material, $M_{Re}$.


\begin{figure}[!htbp]
	\begin{circuitikz}[scale=0.8,american voltages]
		
		\draw
		(0,0) to[esource, name=pulse1, l=$E$] (0,4)
		to[R, l=$R_T$] (3,4)
		to[L, l_=$j\omega L_T$] (3,0)
		-- (0,0);
		\drawpulse{pulse1}
		
		\draw
		(5,2) to[L, l_=$j\omega L_R$] (5,4)
		to[R, l=$R_R$] (8,4) to[generic, l=$Z_L$] (8,2) -- (5,2);
		
		\draw
		(5,0) to[L, l_=$j\omega L_e$] (5,2)
		(5,2) -- (8,2) -- (8,0)
		to[R, l=$R_e$] (5,0);
		
		\draw[->] (7.4,2.8) arc[start angle=0,end angle=285,radius=0.4]
		node[midway,above]{$I_R$};
		\draw[->] (7.4,0.8) arc[start angle=0,end angle=285,radius=0.4]
		node[midway,above]{$I_e$};
		\draw[->] (1.5,1.9) arc[start angle=285,end angle=0,radius=0.5]
		node[midway,above]{$I_T$};
		
		\draw (3.2,2.8) node{$.$};
		\draw (3.2,2.9) node{$.$};
		
		\draw (4.8,3.8) node{$.$};
		\draw (4.8,3.9) node{$.$};
		
		\draw (4.8,1.8) node{$.$};
		\draw (4.8,1.9) node{$.$};
		
		\draw[bend left] (3.3,2.9) to node[midway,above]{$M_{TR}$} (4.7,3.9);
		\draw[bend right] (3.3,2.8) to node[midway,below]{$M_{Te}$} (4.7,1.8);
		\draw[bend right] (4.7,3.8) to node[midway,below]{$M_{Re}$} (4.7,1.9);
		
		
	\end{circuitikz}
	\caption{Linear equivalent circuit of the transmitter coil, receiving coil and the equivalent coil representing eddy currents in Ti.}
	\label{cir_eq1}
\end{figure}
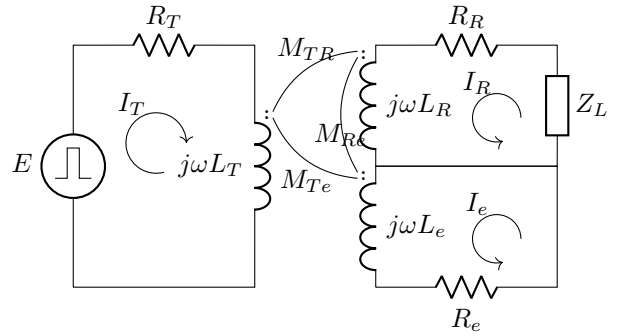

The effects of the magnetic couplings between the three coils in the model can be represented by dependent sources, as illustrated in Fig.~\ref{cir_eq2}.

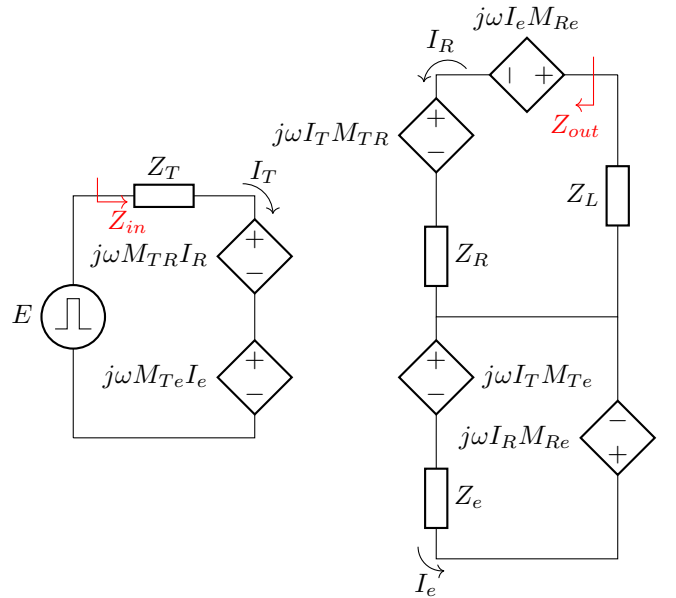
\begin{figure}[!htbp]
	\begin{circuitikz}[scale=0.8,american voltages]
		
		\draw
		(0,0) to[esource, name=pulse1, l=$E$] (0,4)
		to[generic, l=$Z_T$] (3,4)
		to[controlled voltage source, l_=$j\omega M_{TR} I_R$] (3,2.0)
		to[controlled voltage source, l_=$j\omega M_{Te} I_e$] (3,0)
		-- (0,0);
		
		\drawpulse{pulse1}
		
		\draw[->] (2.8,4.2) arc[start angle=90,end angle=0,radius=0.5]
		node[midway,above]{$I_T$};
		
		\draw
		(9,6) to[controlled voltage source, l_=$j\omega I_e M_{Re}$] (6,6) to[controlled voltage source, l_=$j\omega I_T M_{TR}$] (6,4)
		to[generic, l=$Z_R$] (6,2.0) -- (9,2)
		to[generic, l=$Z_L$] (9,6);
		
		\draw
		(6,2) to[controlled voltage source, l=$j\omega I_T M_{Te}$] (6,0)
		to[generic, l=$Z_e$] (6,-2.0) -- (9,-2)
		to[controlled voltage source, l=$j\omega I_R M_{Re}$] (9,2);
		
		\draw[->] (6.5,6.1) arc[start angle=40,end angle=175,radius=0.4]
		node[midway,above]{$I_R$};
		\draw[->] (5.7,-1.8) arc[start angle=180,end angle=275,radius=0.4]
		node[midway,below]{$I_e$};
		
		\draw[red,->] (0.4,4.3) -- ++(0,-0.4) -- ++(0.5,-0.0)  node[below]{$Z_{in}$};
		\draw[red,->] (8.6,6.3) -- ++(0,-0.8) -- ++(-0.3,-0.0) node[below]{$Z_{out}$};
		
	\end{circuitikz}
	\caption{Linear equivalent circuit of the coils with dependent sources representing induced voltages.}
	\label{cir_eq2}
\end{figure}


\begin{figure*}[!t]
	\begin{equation}
	Z_{in}=\frac{Z_T(Z_R+Z_L)Z_e+\omega^2M_{Re}^2Z_T+\omega^2M_{TR}^2Z_e+\omega^2M_{Te}^2(Z_R+Z_L) -2j(\omega^3 M_{Re} M_{TR} M_{Te} )}{(Z_R +Z_L)(Z_T)+(\omega^2 M_{Re}^2)}
	\label{zin}
	\end{equation}
\end{figure*}
\begin{figure*}[!t]
	\begin{equation}
	Z_{out}=\frac{Z_T Z_R Z_e+Z_T\omega^2 M_{Re}^2+\omega^2 M_{TR}^2 Z_{e} +\omega^2 Z_{R} M_{Te}^2-2j\omega^3 M_{TR} M_{TR} M_{Re}}{Z_T Z_e+\omega^2 M_{Te} ^2} 
	\label{zout}
	\end{equation}
\end{figure*}

Eddy currents influence the input impedance as shown in (\ref{zin}) and the output impedance as shown in (\ref{zout}). Considering the circuit in Fig.~\ref{cir_eq2} with the parameters given in Table~\ref{tab_1} at a frequency of 350 kHz, Figs.~\ref{fig_zre}, \ref{fig_zle}, and \ref{fig_zmre} illustrate the effect of varying the values of the parameters $R_e$ (resistor), $L_e$ (inductor), and $M_{Re}$ (mutual inductance between the receiving circuit and the induced current paths) on the input impedance $Z_{in}$ and output impedance $Z_{out}$.

\begin{table}[]
	\caption{Parameters used in the simulation of the Linear Equivalent Circuit -- close to the values of the parameters used in the validation procedure.}
	
	\begin{center}
	\begin{tabular}{llll}
		\hline
		Parameter & $R_T$         & $R_R$       & $R_e$       \\
		Value     & 3 $\Omega$    & 5 $\Omega$  & 5 $\Omega$  \\ \hline
		Parameter & $L_T$         & $L_R$       & $L_e$       \\
		Value     & 1765.9 $\mu$H & 28 $\mu$H   & 1 $\mu$H    \\
		\hline
		Parameter & $M_{TR}$      & $M_{Te}$    & $M_{Re}$    \\
		Value     & 1.1 $\mu$H    & 0.21 $\mu$H & 0.03 $\mu$H \\
		\hline
	\end{tabular}
\end{center}
	\label{tab_1}
\end{table}

\begin{figure}[!htbp]
	\centerline{\includegraphics[width=1.0\columnwidth]{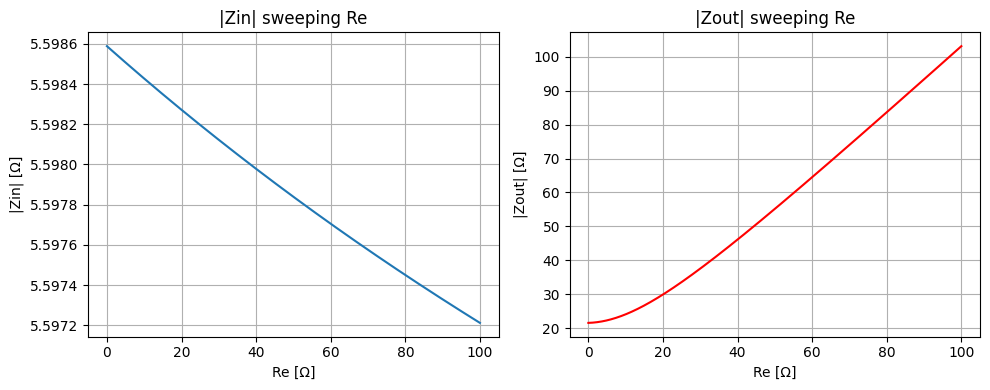}}
	\caption{$Z_{in}$ and $Z_{out}$ keeping fixed parameters while sweeping $R_e$ form 0 to 100 $\Omega$.}
	\label{fig_zre}
\end{figure}  
\begin{figure}[!htbp]
	\centerline{\includegraphics[width=1.0\columnwidth]{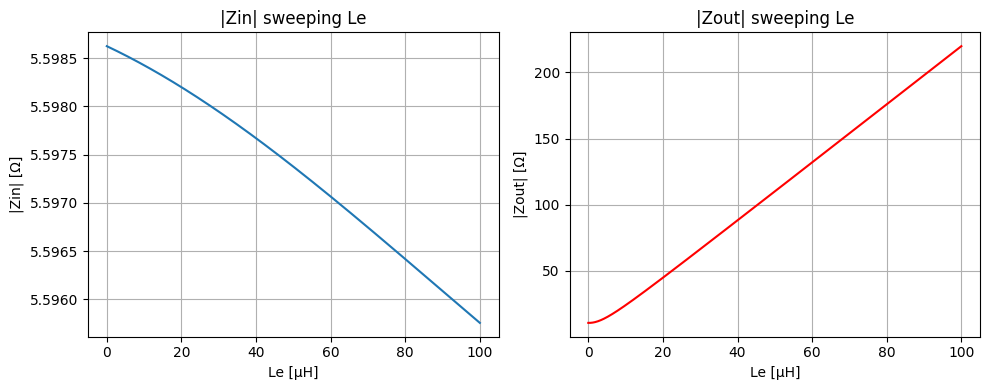}}
	\caption{$Z_{in}$ and $Z_{out}$ keeping fixed parameters and sweeping $L_e$ from 0 to 100 $\mu$H.}
	\label{fig_zle}
\end{figure}  
\begin{figure}[!htbp]
	\centerline{\includegraphics[width=1.0\columnwidth]{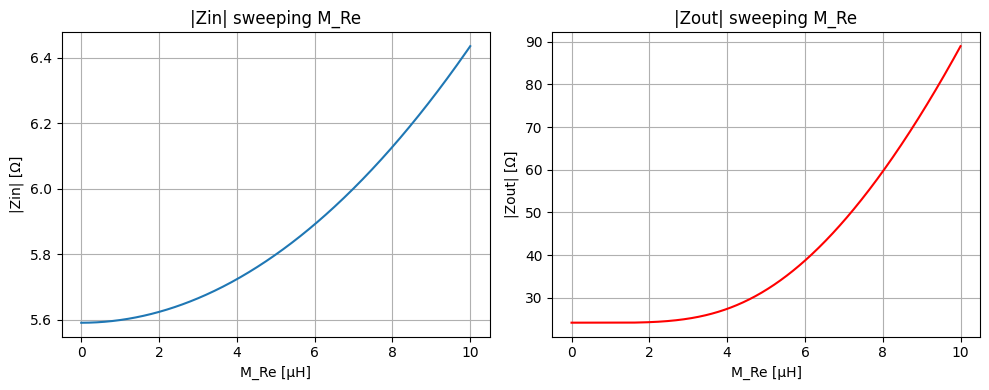}}
	\caption{$Z_{in}$ and $Z_{out}$ keeping fixed parameters and sweeping $M_Re$ from 0 to 10 $\mu$H.}
	\label{fig_zmre}
\end{figure}

These figures show that the impact of eddy currents is particularly significant on the output impedance $Z_{out}$. In addition to the energy loss dissipated in $R_e$ and the attenuation of the effective magnetic flux in the receiving coil, the impedance change inevitably leads to a shift in the resonance frequency.

\section{Cavity in a Conductive Implant}

Fig.~\ref{protese1}(a) illustrates a simplified generic structure consisting of four Ti walls ($ 17 \times 20 \times 10$ mm and $0.5$ mm thickness) enclosing a solenoidal receiving coil. The receiving coil, together with the associated electronics (rectifier and regulator), can be positioned inside the walls, forming a cavity. Fig.~\ref{protese1}(b) shows the simulation results for a magnetic field applied by a transmitting coil separated by a distance of $8.9$ cm of the receiving coil. The transmitting coil consists of two sections solenoids with a diameter of $\diameter = 15$ cm, spaced $17.8$ cm apart, with a total inductance of $L_1 = 1.7659$ mH. The receiving coil has a diameter of $\diameter = 9.1$ mm (with high permeability core $\diameter = 8.5$ mm and $5.5$ mm height) and an inductance of $L_2 = 28.0$ $\mu$H. These parameters are listed in Table~\ref{tab_1} and these coils will be used in the whole work. It can be observed that the parasitic (eddy) currents generated in the Ti walls almost completely cancel the magnetic fields inside the cavity, thereby drastically reducing the energy transmitted to the receiving circuit. It is also evident that the induced currents in the conductor cause energy dissipation, leading to increased losses and reduced efficiency.

\begin{figure}[!htbp]
	\centerline{\includegraphics[width=1.0\columnwidth]{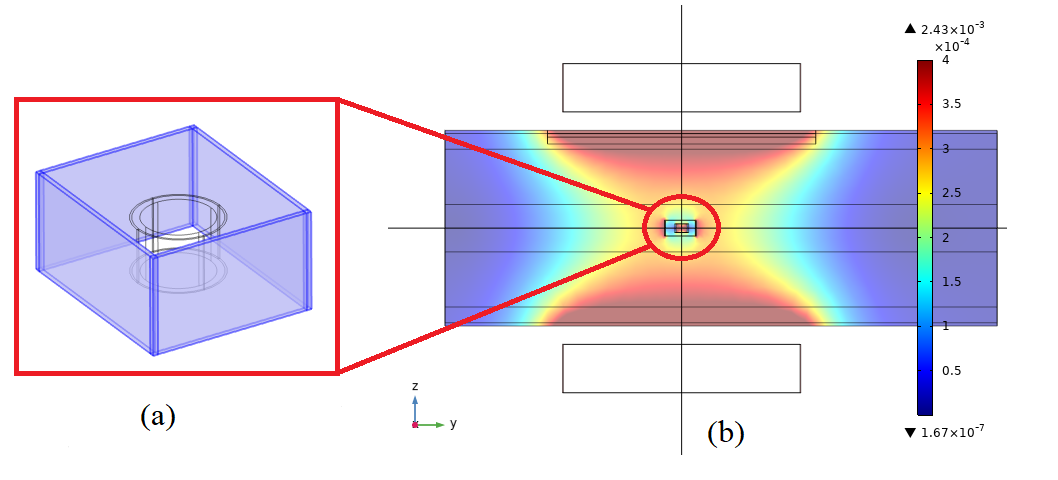}}
	\caption{(a) Generic representation of a cavity with four walls ($ 17 \times 20 \times 10$ mm and $0.5$ mm thickness) in a Ti conductor designed to house the receiving coil and rectifier circuit; (b) Simulation of the magnetic field [T] distribution in the structure when powered by a two-section solenoid transmitting coil.}
	\label{protese1}
\end{figure}

This geometry creates a closed path for the induced currents, whose intensity depends on factors such as the total perpendicular magnetic flux, excitation frequency, and material conductivity. The dimensions of the cavity walls, as well as their thicknesses, have a significant influence on the magnetic coupling factor $k$ between an external transmitting coil and a receiving coil. Fig.~\ref{espessura} illustrates the effect of the Ti-6Al-4V wall thickness on the magnetic coupling factor $k$ between the transmitting coil and the receiving coil.

\begin{figure}[!htbp]
	\centerline{\includegraphics[width=1.0\columnwidth]{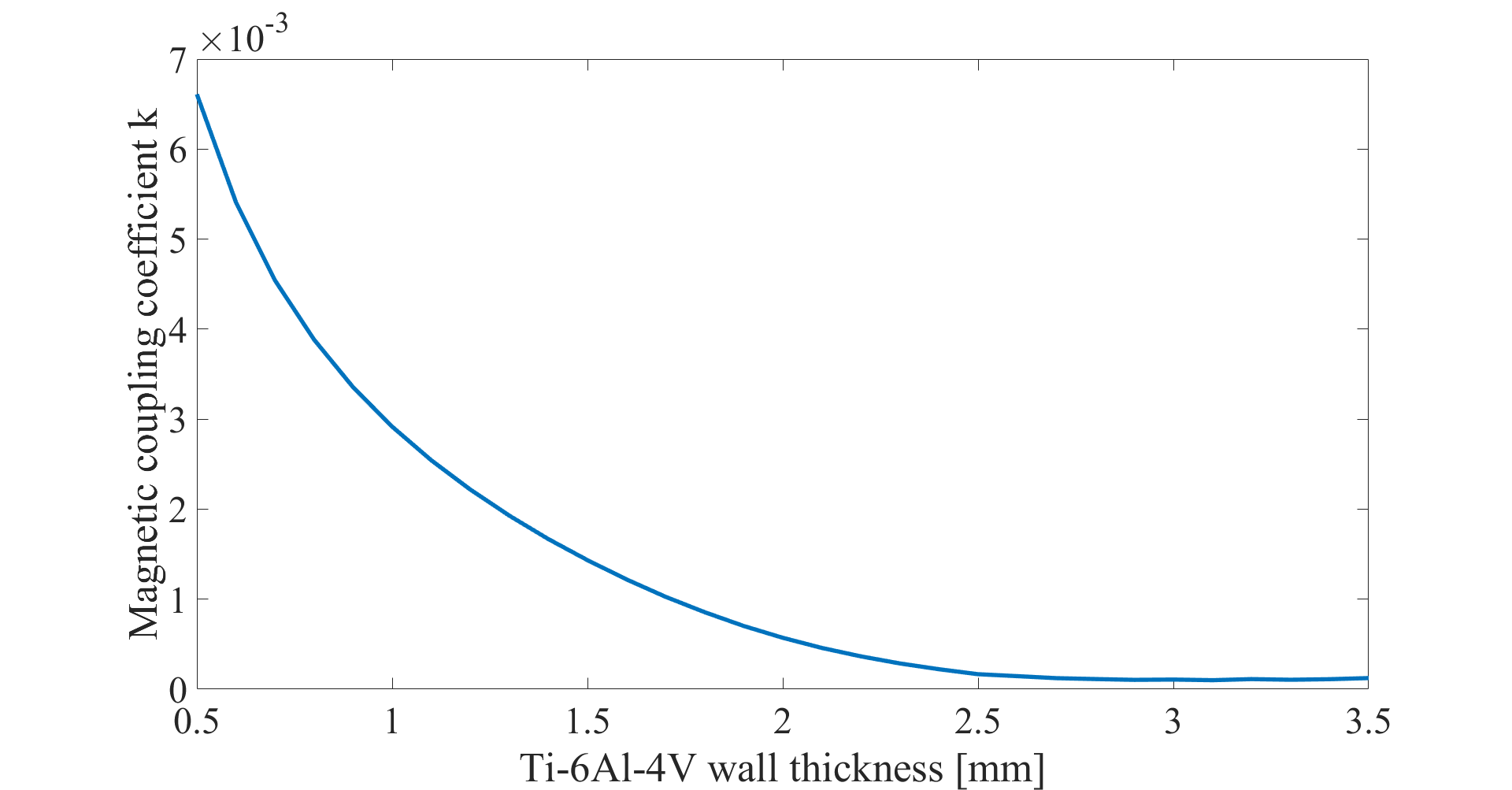}}
	\caption{Influence of the Ti wall thickness on the magnetic coupling factor $k$ between the transmitting and receiving coils (Fig.\ref{protese1}).}
	
	\label{espessura}
\end{figure}

\subsection{Mitigation of the Effect of Induced Currents in the Cavity Structure for the Solenoid}

A possible solution to mitigate this effect is to increase the magnetic reluctance of the current path by reducing the wall thickness or the overall volume of conductive material, thereby decreasing the eddy current density $J_{eddy}$. Fig.~\ref{protese2} shows the reduction of the conductor volume achieved by introducing transverse holes in the wall of the structure shown in Fig.~\ref{protese1}(a). The direct consequence of this structural modification is an increase in the magnetic coupling coefficient $k$ between the transmitting and receiving coils (Table~\ref{Tab_k}). This means that a larger fraction of the magnetic field generated by the transmitting coil is effectively captured by the receiving coil, improving the overall system performance.

\begin{figure}[!htbp]
	\centerline{\includegraphics[width=.8\columnwidth]{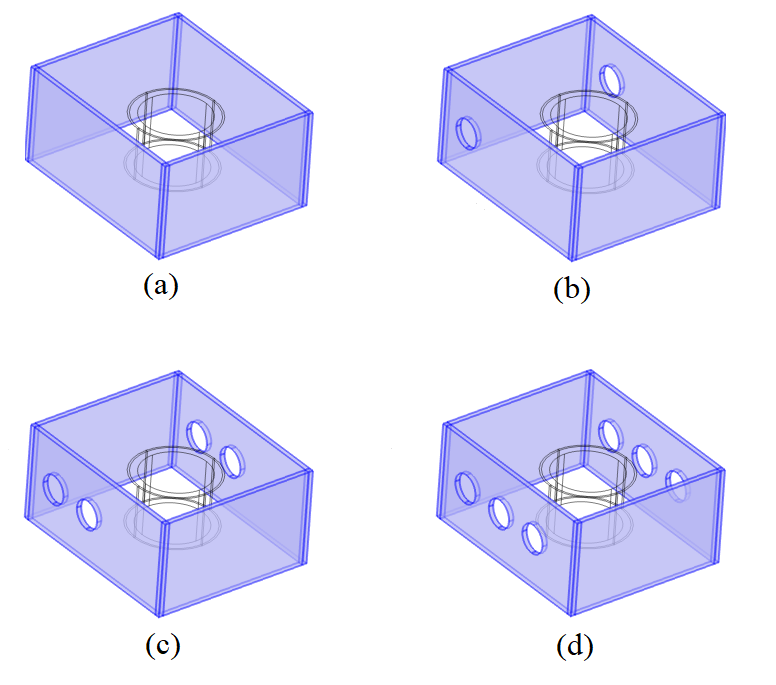}}
	\caption{(a) Solid structure; (b) Structure with two lateral holes of $\diameter = 3$ mm; (c) Structure with four lateral holes of $\diameter = 3$ mm; (d) Structure with six lateral holes of $\diameter = 3$ mm.}
	\label{protese2}
\end{figure}
This effect can be further enhanced by completely interrupting the conductive path around the receiving coil, as illustrated in Fig.~\ref{protese3}. In this case, the induced currents have a significantly reduced influence on the magnetic flux linking the transmitting and receiving coils. Fig.~\ref{protese4} compares the magnetic flux provided by the transmitting coil for: (a) a solid structure and (b) the structure with the eddy currents loop interrupted.

\begin{figure}[!htbp]
	\centerline{\includegraphics[width=.5\columnwidth]{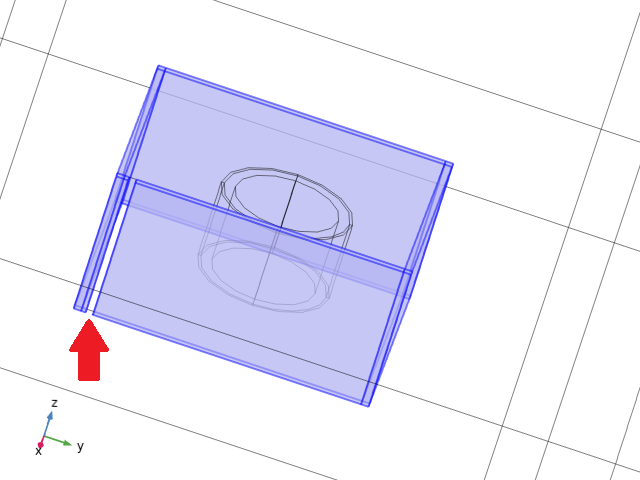}}
	\caption{Interruption of the eddy current path generated in the Ti structure eliminating the conductive loop.}
	\label{protese3}
\end{figure}

\begin{figure}[!htbp]
	\centerline{\includegraphics[width=1\columnwidth]{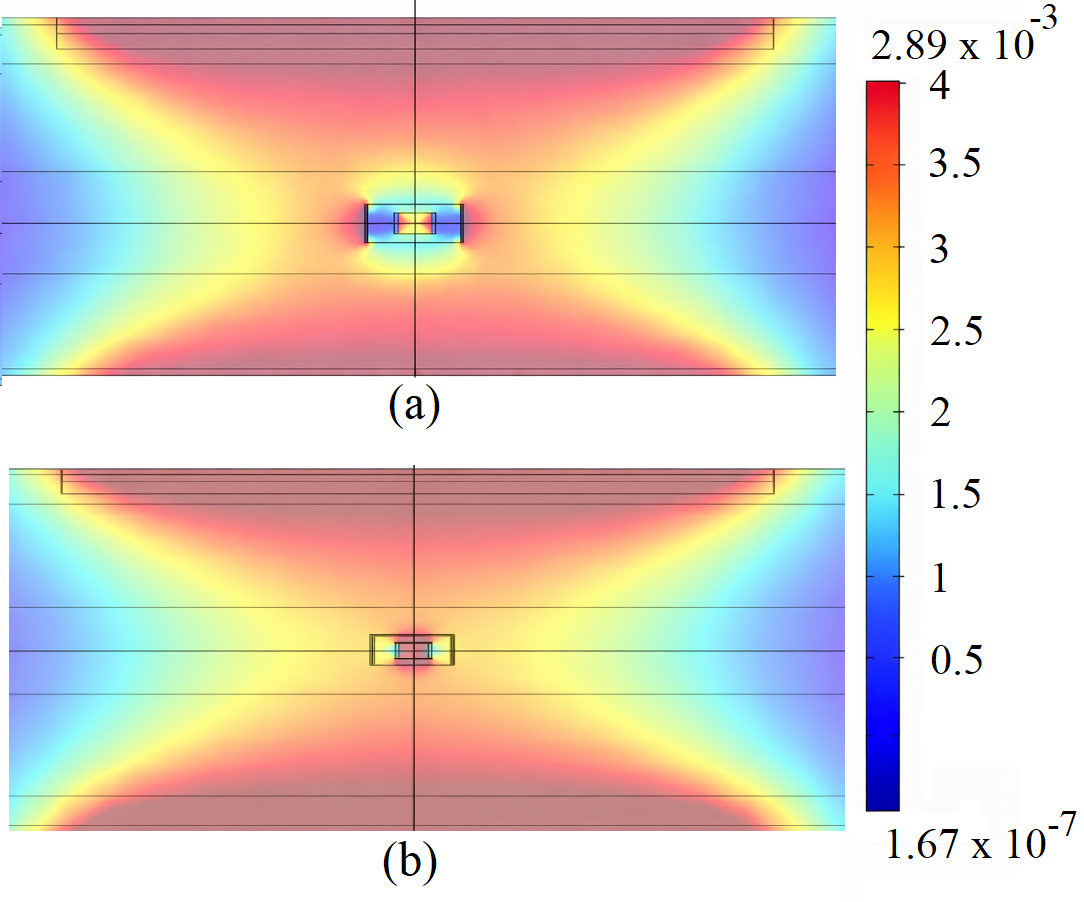}}
	\caption{Magnetic Field [T] around the Ti structure: (a) with eddy currents loops; (b) with interrupted eddy currents path (Fig.~\ref{protese3}).}
	\label{protese4}
\end{figure}

It is also possible to consider the use of shielding with a material of high magnetic permeability inside the cavity. With this shielding, the magnetic flux tends to concentrate along the path of least reluctance, thereby reducing the effect of induced currents. Fig. \ref{blindagem} shows the effect of shielding with the material \textbf{IFL16} (relative permeability = 220 up to 3 Mhz) on the previously presented closed structure for comparison purposes. It can be observed that the magnetic flux concentrates near the walls of the Ti structure.

\begin{figure}[!htbp]
	\centerline{\includegraphics[width=.6\columnwidth]{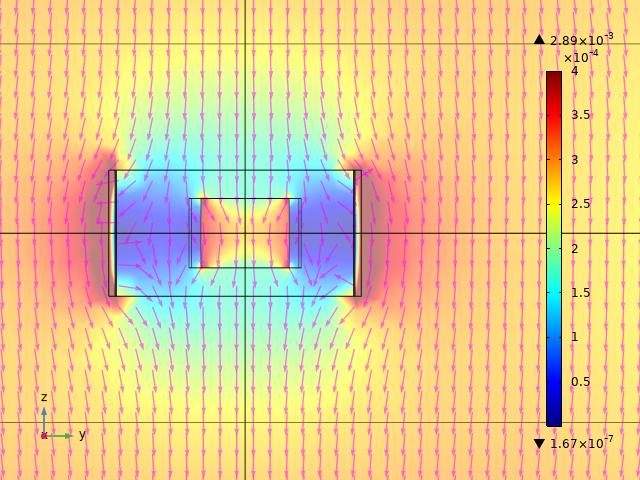}}
	\caption{ Effect on the distributed magnetic field [T] caused by the shielding implemented by a high-permeability film (\textbf{IFL16} - relative permeability = 220 up to 3 Mhz) on the inner walls of the cavity, near the coil.}
	\label{blindagem}
\end{figure}

The concept of the four-wall structure can be applied to the design of a prosthesis, whose geometry must be adapted according to the intended application. As an illustrative example, in this work we consider a solid Ti bar with one end containing the four walls designed to accommodate the receiving coil, as shown in Fig.~\ref{protese}. The total dimensions of this structure is $17\times 80 \times10$ mm. Fig.~\ref{protese} illustrates the magnetic flux density [T] distribution, when the path of the eddy currents is interrupted, allowing the transmitting coil to link with the receiving coil, thereby enabling the transfer of a larger amount of energy with reduced losses.

\begin{figure}[!htbp]
	\centerline{\includegraphics[width=1\columnwidth]{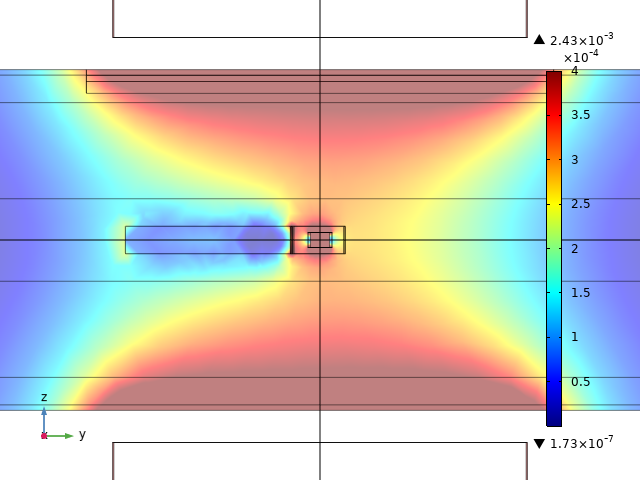}}
	\caption{Hypothetical prosthesis constructed with a solid body [Ti] and an end featuring a slot for fitting the ferrite-core receiver coil \diameter $9.5$ mm $\times 5.5$ mm [T].}
	\label{protese}
\end{figure}

Table~\ref{Tab_k} presents the values of the magnetic coupling coefficients $k$ evaluated both for the simplified structures of Fig.\ref{protese1} and \ref{protese}. This table illustrates the inductive coupling factor $k$ between transmitting and receiving coils for all the structural modifications analyzed: the inclusion of holes in the walls, the use of a shielding film inside the cavity and the total interruption of eddy currents path. This latter condition is simulated with the receiving coil both with (f) and without (g) a magnetic core.


\begin{table}[h]
	\caption{Magnetic Coupling Coefficient of simulated cases.}
	\centering
	\begin{tabular}{l|l|l|}
		\cline{2-3}
		& \begin{tabular}[c]{@{}l@{}}4 solid \\ walls k\end{tabular} & \begin{tabular}[c]{@{}l@{}}Hypothetical\\ prosthesis k\end{tabular} \\ \hline
		\multicolumn{1}{|l|}{Solid structure - Fig.~\ref{protese2} (a)}                                                             & 0.0066094                                                  & 0.0050785                                                           \\ \hline
		\multicolumn{1}{|l|}{2 holes - Fig~\ref{protese2} (b)}                                                                      & 0.0069475                                                  & 0.0054201                                                           \\ \hline
		\multicolumn{1}{|l|}{4 holes - Fig~\ref{protese2} (c)}                                                                      & 0.0073371                                                  & 0.0058555                                                          \\ \hline
		\multicolumn{1}{|l|}{6 holes - Fig.~\ref{protese2} (d)}                                                                     & 0.0077127                                                  & 0.0062778                                                           \\ \hline
		\multicolumn{1}{|l|}{shielding - Fig.~\ref{blindagem}}                                                                   & 0.0071982                                                  & 0.0065045                                                           \\ \hline
		\multicolumn{1}{|l|}{\begin{tabular}[c]{@{}l@{}}Loop Interruption-coil\\ with core - Fig.~\ref{protese3}  \end{tabular}}   & 0.024061                                                   & 0.025967                                                            \\ \hline
		\multicolumn{1}{|l|}{\begin{tabular}[c]{@{}l@{}}Loop Interruption-coil\\ without core \end{tabular}} & 0.010943                                                   & 0.011795                                                            \\ \hline
	\end{tabular}
	\label{Tab_k}
\end{table}

The numerical results show that the modification with the greatest impact on the magnetic coupling coefficient $k$ is the interruption of the path of the eddy currents around the receiver coil. For the parameters used, it is observed that with the closed eddy currents loop, $k = 0.0066094$, and after introducing the slit, this value increases to $k = 0.025114$ (at 340 kHz). This increase represents a significant improvement in system efficiency, indicating that a much larger fraction of the magnetic flux generated by the transmitter coil links to the receiver coil.

\subsection{Planar Coils}
The geometry of the structure plays a decisive role in the performance of a wireless power transfer system, especially in cases involving implants made of conductive materials, as illustrated in the presented examples.

An alternative to the cavity-based configuration is the use of planar coils, which offer easy mounting and can be shielded on one of their surfaces, thereby reducing electromagnetic flux leakage into the solid Ti body. Fig.~\ref{fig:bob_planar}(a) shows a planar spiral coil over a IFL16 ferrite sheet, whose function is to enhance magnetic coupling and minimize losses. Fig.~\ref{fig:bob_planar}(b) shows the planar spiral coil attached to the Ti surface.





\begin{figure}[!htbp]
	\centerline{\includegraphics[width=.7\columnwidth]{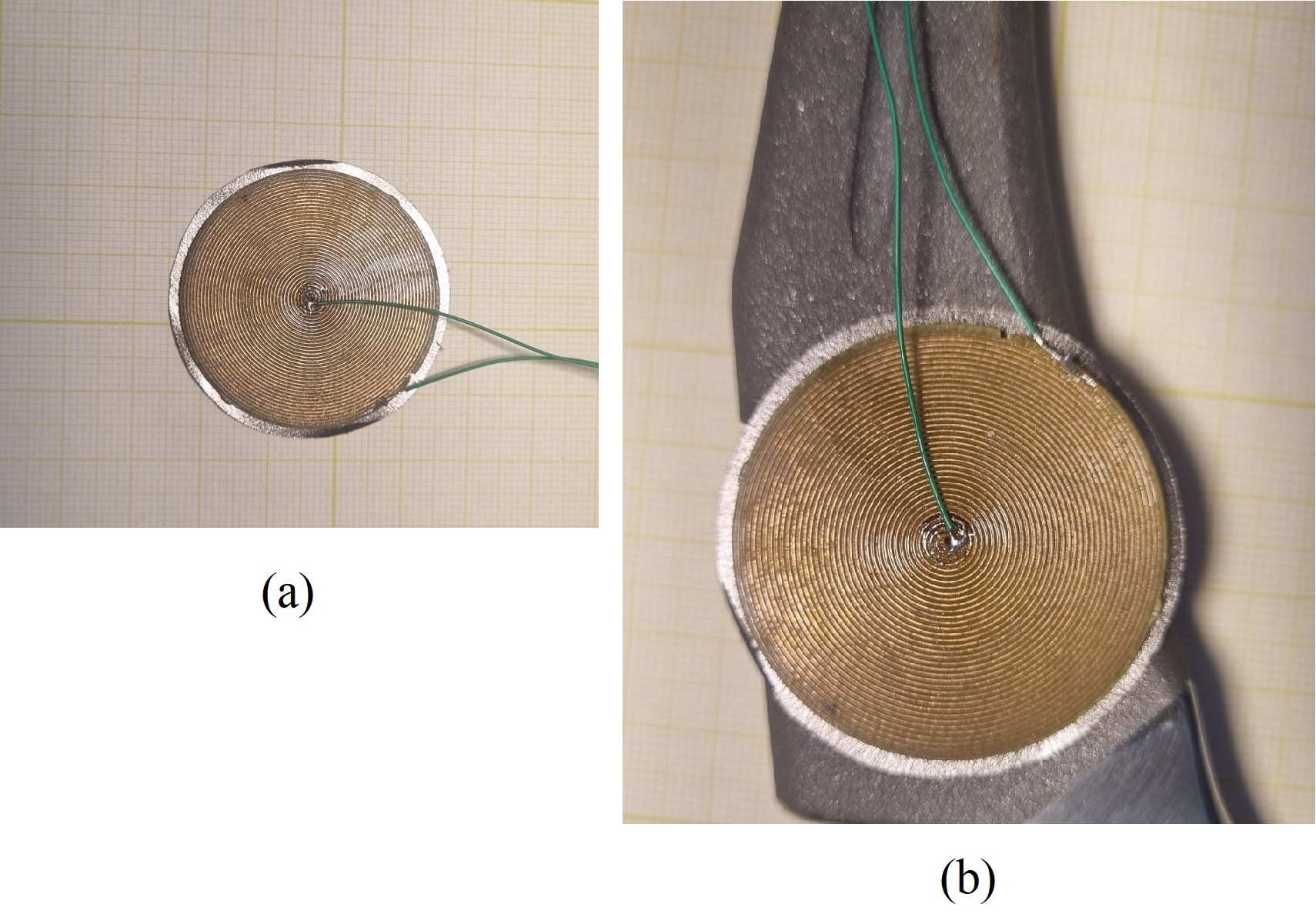}}
	\caption{(a) Planar spiral coil over a layer of high-permeability material (ferrite sheet). (b) Planar spiral coil attached to the Ti surface.}
	\label{fig:bob_planar}
\end{figure}

\section{Experimental Setup and Results}



A WPT system based on magnetic coupling was implemented to deliver energy to a coil embedded in a Ti structure, with the aim of experimentally validating the simulated conditions. The transmitting coil comprises two solenoids connected in series, each with a diameter of $\diameter = 150$ mm, a width of $w = 20$ mm, and $N = 10$ turns. The solenoids are separated by a distance of $d = 180$ mm, as illustrated in Fig.~\ref{fig:estrutura}. The transmitting coil assembly presents an inductance of $L_1 = 58.2~\mu$H and a series equivalent resistance (ESR) of $R_1 = 0.48~\si{\ohm}$.

\begin{figure}[h!]
	\begin{center}
		\includegraphics[width=0.9 \columnwidth]{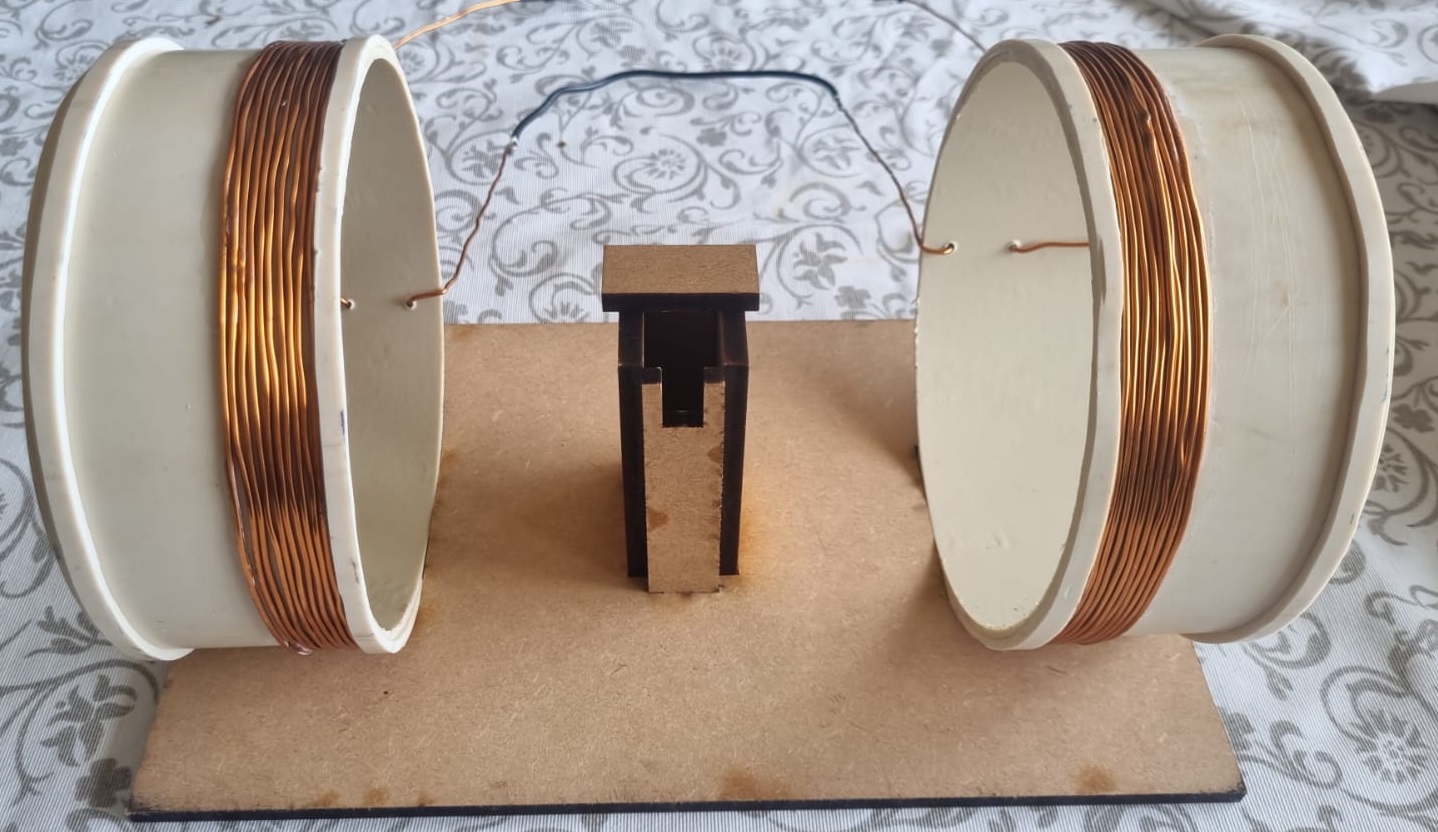}
	\end{center}
	\caption{Transmitting coil set.}
	\label{fig:estrutura}
\end{figure}

The dimensions of the receiving coil are constrained by the available space inside the cavity of the implant. Figure~\ref{fig:receiving}(a) shows the implant cavity used in this work, whose approximate dimensions are 9 $\times$ 22 $\times$ 5~mm. Accordingly, a winding with $N = 30$ turns of AWG~26 enameled wire was fabricated, occupying approximate dimensions of 8 $\times$ 16 $\times$ 4~mm.
The core employed was a ferrite bar with a rectangular cross section of 13 $\times$ 5~mm and thickness of 4~mm. 
This ferrite core operates in an open magnetic path configuration, resulting in an effective relative permeability of approximately $\mu_{\mathrm{eff}} \approx 1.25$, as inferred from the measured inductance. Additionally, a ferrite shielding sheet IFL16 was applied along the entire inner wall of the cavity. The receiving coil embedded in the cavity exhibits an inductance of $L_2 = 23.0~\mu$H and a ESR of $R_2 = 1.6~\si{\ohm}$.

\begin{figure}[h!]
	\begin{center}
		\includegraphics[width=1.0 \columnwidth]{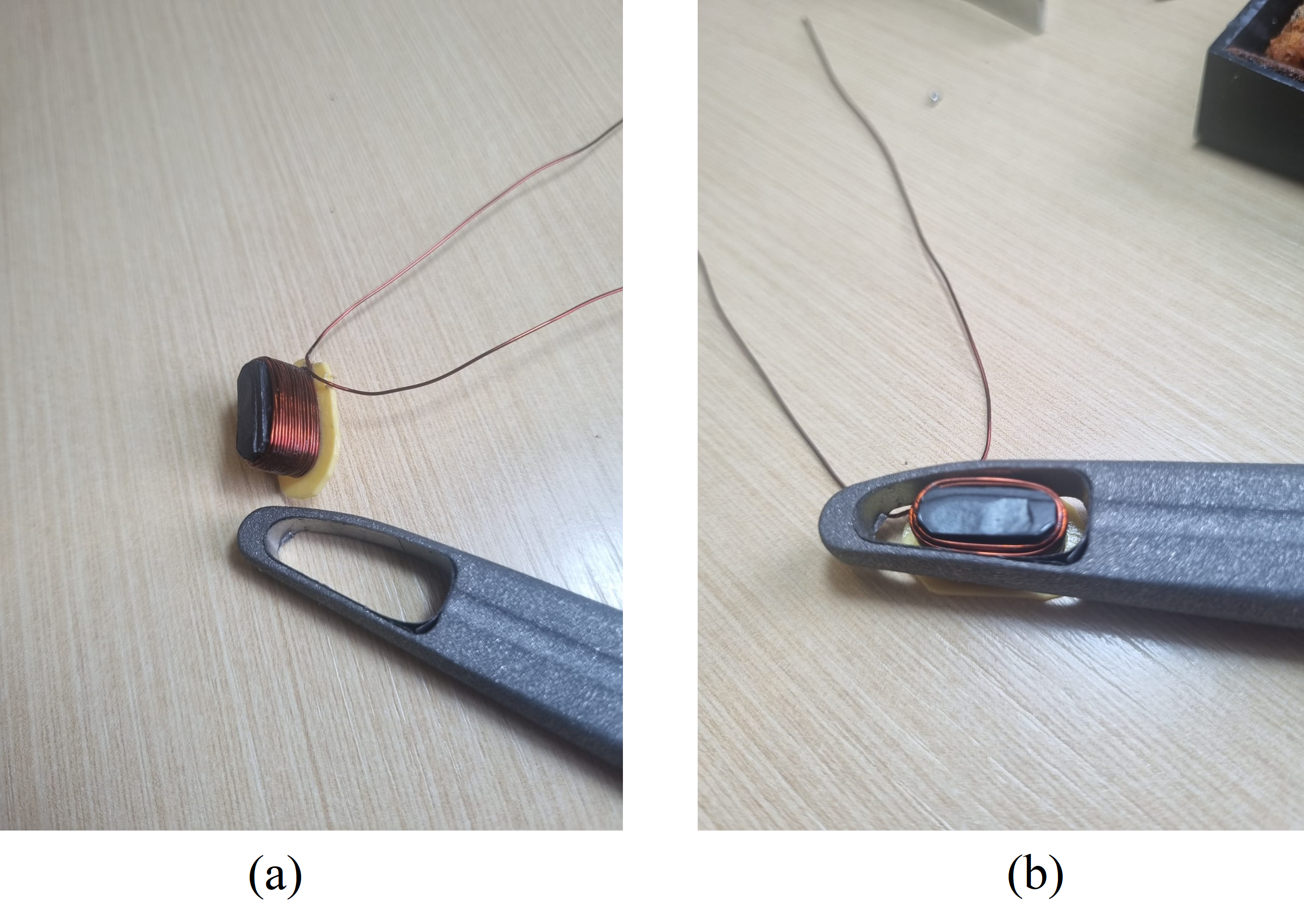}
	\end{center}
	\caption{Receiving coil: (a) separated parts, and (b) embedded in the Ti implant.}
	\label{fig:receiving}
\end{figure}

\subsection{Magnetic coupling coefficient estimation}

The magnetic coupling coefficient between the transmitting and receiving coils was estimated from the open-circuit voltage measurement across the receiving coil ($L_2$), denoted as $V_{2,\mathrm{oc}}$ and the corresponding sinusoidal current in the transmitting coil $I_1$, at the resonance frequency $f$. Under these conditions, the mutual inductance $M$ can be defined as: 
\begin{equation}
M=\frac{V_{2,\mathrm{oc}}}{\omega I_1} 
\label{eq:metodoA}
\end{equation}
where $M=k \sqrt{L_1 L_2}$, $\omega = 2\pi f$, $L_2 = 23~\mu\text{H}$, and $L_1=58.2~\mu\text{H}$. When the receiver coil is positioned at the center of the two sections of the transmitting coil, which are separated by a distance of ( $d = 18$ ~\text{cm} ) (thus resulting in a 9 cm separation between the transmitting and receiving coils), the magnetic coupling coefficient is $k = 0.002$ at $f = \SI{273}{\kilo \hertz}$.

\subsection{Capacitive Compensation Network}

There are several matching network topologies reported in the literature~\cite{Azambuja2014}. In this work, an S--P topology, consisting of series compensation at the transmitter ($C_1$) and parallel ($C_2$) compensation at the receiver, was adopted. Since the transmitter-side capacitor is subjected to high voltage stress, a capacitor array consisting of four series-connected capacitors per branch and four parallel branches was implemented. Multilayer ceramic capacitors rated at $5.6~\si{\nano\farad}/1~\si{\kilo\volt}$ were used, resulting in an equivalent capacitance of $C_1 = 5.78~\si{\nano\farad}$ and an equivalent series resistance (ESR) of $R_{\mathrm{ESR},c} = 1.6~\si{\ohm}$.

Series compensation at the transmitter input maximizes the current and, consequently, the generated magnetic field. The values of the compensation capacitors $C_1$ and $C_2$, together with the operating frequency $f$, allow multiple solutions for maximizing the power delivered to the load $R_L$. By fixing $C_1$ at $5.78~\si{\nano\farad}$, the number of feasible solutions is reduced, constraining the operating frequency to a region close to $300~\si{\kilo\hertz}$. The determination of the operating frequency $f$ and the value of the receiver-side capacitor $C_2$ can therefore be efficiently carried out through an optimization procedure, whose objective function is the maximization of the power delivered to the load $R_L$. Accordingly, simulation results yield $C_2 = \SI{14.6}{\nano\farad}$ and $f = \SI{265}{\kilo\hertz}$.

Parallel compensation at the receiver maximizes the induced output voltage. A parallel capacitor of $C_2 = 15~\si{\nano\farad}$ was implemented to complete the matching network and tune the receiving circuit.

\subsection{Transmitter and Receiver Electronic Circuits} 

The WPT circuit is powered by an electronic inverter implemented with a FET-based bridge capable of operating at frequencies up to \SI{300}{\kilo\hertz}. Fig.~\ref{fig:WPT_circuit} shows the simplified electronic circuit, where the gate signals $g_1$, $g_2$, $g_3$, and $g_4$ are generated by the IR2104 integrated gate driver.

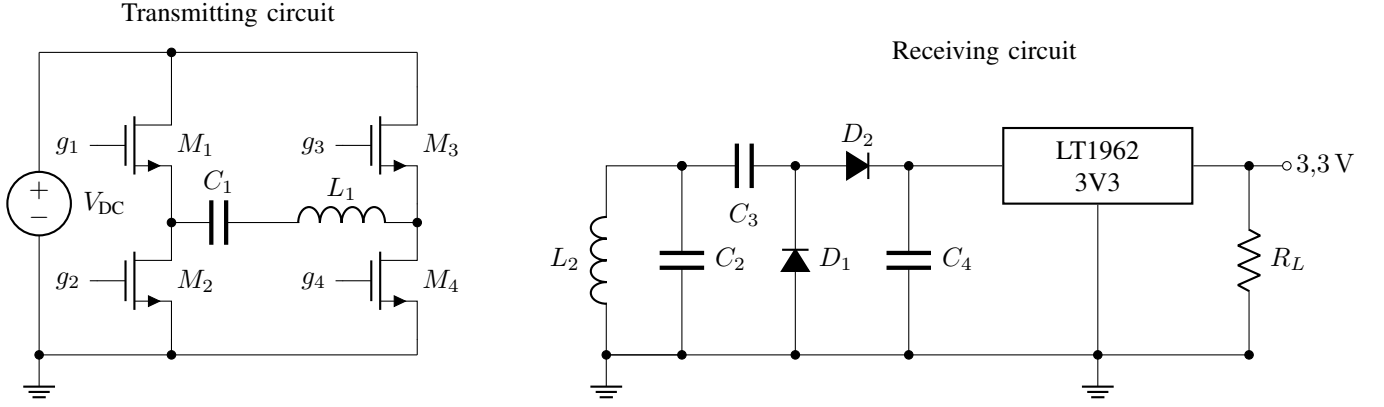
\begin{figure*}[!htbp]
	\begin{circuitikz}[scale=0.5,american voltages]
		\ctikzset{tripoles/mos style/arrows}
		\draw
		(0,0) node[ground]{} to[V, l_=$V_{\text{DC}}$, invert] (0,8);
		\draw
		(0,8) to[short] (10,8);
		\draw
		(0,0) to[short,*-] (10,0);
		
		\draw
		(3.5,4) node[nmos, anchor=S] (M1) {$M_1$}			
		(M1.D) -- (3.5,8)   
		(M1.S) -- (3.5,3.5);     
		\draw (3.5,8) node[circ]{};				
		\draw
		(3.5,3.5) node[nmos, anchor=D] (M2) {$M_2$}
		(M2.S) -- (3.5,0)      
		(M2.D) -- (3.5,3.5);     
		\draw (3.5,0) node[circ]{};				
		\draw
		(10,4) node[nmos, anchor=S] (M3) {$M_3$}
		(M3.D) -- (10,8)      
		(M3.S) -- (10,3.5);     
		
		\draw
		(10,3.5) node[nmos, anchor=D] (M4) {$M_4$}
		(M4.S) -- (10,0)      
		(M4.D) -- (10,3.5);     
		
		\draw
		(M1.G) -- +( -0.2,0) node[left]{$g_1$};
		\draw
		(M2.G) -- +( -0.2,0) node[left]{$g_2$};
		\draw
		(M3.G) -- +(  -0.2,0) node[left]{$g_3$};
		\draw
		(M4.G) -- +(  -0.2,0) node[left]{$g_4$};
		
		
		\draw
		(3.5,3.5) to[C, l=$C_1$] (6,3.5) to[L, l=$L_1$] (10,3.5);
		\draw (3.5,3.5) node[circ]{};
		\draw (10,3.5) node[circ]{};	
		
		\draw
		(15,0) node[ground]{};
		
		\draw
		(15,0) to[L, l=$L_2$] (15,5) to [short, -*] (17,5) to [C, l=$C_2$] (17,0)
		to[short, *-] (15,0);

		
		
		\draw
		(17,5) to [short, *-] (17.3,5) to[C,  l_=$C_3$] (20,5);
		\draw
		(15,0) to [short, *-] (20,0) to[D*, l_=$D_1$] (20,5);
		
		\draw
		(20,5) to [short, *-] (20.3,5) to[D*, l=$D_2$] (23,5)
		to[C,  l=$C_4$] (23,0) to [short, -*] (20,0);
		
		
		\node[draw, thick,
		minimum width=2.5cm,
		minimum height=1.0cm,
		align=center]
		(reg) at (28,5)
		{LT1962\\3V3};
		
		\draw
		(23,5) to[short, *-] (reg.west);
		\draw
		(reg.south) to[short] (28,0) to [short, -*] (23,0);
		
		\draw
		(reg.east) to[short,-o] (33,5) node[right]{$3{,}3\,\mathrm{V}$};
		
		\draw
		(32,5) to[R, l=$R_L$] (32,0)
		to[short] (32,0);
		\draw (32,0) to [short,*-*] (28,0) node[ground]{};
		\draw (32,5) node[circ]{};
		\node at (5,9) {Transmitting circuit};
		\node at (25,8) {Receiving circuit};

	\end{circuitikz}
	\caption{WPT proposed circuit. The multiple mutual inductances were omitted for clarity.}
	\label{fig:WPT_circuit}
\end{figure*}
	
In the receiver circuit, a voltage doubler is required because the voltage levels across the parallel LC network ($L_2$ and $C_2$) may be insufficient to adequately bias the voltage regulator integrated circuit. Another important aspect that must be considered is the selection of the rectifying diode. In this application, the Schottky diode ZHCS400 was employed, as it exhibits a typical forward voltage drop of $0.2~\si{\volt}$ and a reverse recovery time of $10~\si{\nano\second}$. In general, the diodes used in rectification stages should present low forward voltage drops and fast switching characteristics, enabling proper operation within the intended frequency range. Finally, the LT1962-3.3 integrated circuit is employed to regulate the output voltage to $3.3~\si{\volt}$, providing an output current of up to $300~\si{\milli\ampere}$ with a low dropout voltage.

\subsection{Experimental Evaluation}

The proposed WPT system was initially evaluated using a solenoid-type receiver embedded in the Ti implant, as shown in Fig.~\ref{fig:receiving}(b). The receiver circuit was inserted into a slot measuring $9$ $\times$ $23$ $\times$ $6$~mm, and the implant was positioned at the midpoint between the transmitting coils, at a distance of 9~cm from each coil. The power supply $V_{DC}$ of the transmitting circuit was set to 12~V, and the load was chosen as $R_L = \SI{3.3}{\kilo\ohm}$ in order to limit the output current to 1~mA. Under these experimental conditions, the experimental optimal frequency was found to be $f = \SI{273}{\kilo\hertz}$. The measured average input current at the transmitting circuit was $I_1 = \SI{2.8}{\ampere}$, resulting in an average input power of $P_i = \SI{34}{\watt}$. On the receiving side, the load voltage was measured as $\SI{1.4}{\volt}$, indicating insufficient level to properly bias the LT1962-3.3 voltage regulator. The power transfer efficiency under these conditions was extremely low, resulting in $\eta = 0.002\%$.


Subsequently, the same experimental setup was evaluated using the Ti implant with the main eddy-current loop interrupted. In this case, the measured average current and input power were 0.59~A and approximately 0.15~W, respectively, confirming that the influence of induced eddy currents in the conductive material can severely compromise system performance when an inadequate geometry is employed. Figure~\ref{experimento} illustrates: (a) the experimental setup comprising the transmitting coils and the prosthesis positioned at the center, with the embedded receiving coil; (b) a detailed view of the receiver coil fitted into the prosthesis; and (c) a close-up view of the slot introduced in the prosthesis to interrupt the eddy-current loop. Under these conditions, the voltage regulator was properly biased, resulting in an output power of $P_o = \SI{3.3}{\milli\watt}$ and a power transfer efficiency of $\eta = 2.2\%$.

\begin{figure}[!htbp]
	\centerline{\includegraphics[width=0.8\columnwidth]{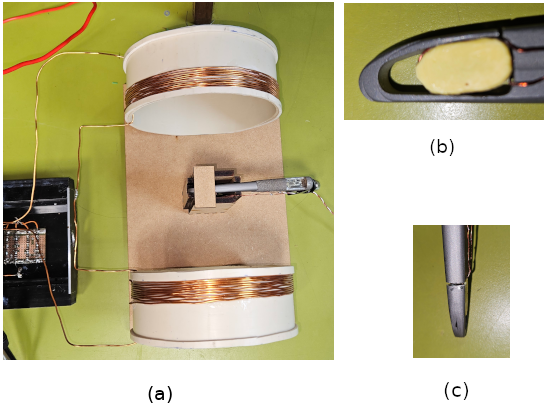}}
	\caption{Experimental setup: (a) the implant positioned at the midpoint between the transmitting coils; (b) a detailed view of the receiving coil embedded in the cavity; and (c) a close-up view of the cut introduced in the structure to interrupt the main eddy-current loop.}
	
	\label{experimento}
\end{figure}

Lastly, a planar spiral receiving coil was used to evaluate the power transfer performance over a Ti implant structure. Figure~\ref{fig:bob_planar} shows the receiving coil employed in this experiment, whose parameters are: diameter $\diameter = \SI{37}{\milli\metre}$, $N = 34$ turns, $L = \SI{21}{\micro\henry}$, and $\mathrm{ESR} = \SI{7.0}{\ohm}$. The experimental setup was the same as that used for the solenoid-type receiving coil. The measured magnetic coupling coefficient was $k = 0.01$, which is similar to the value obtained for the interrupted Ti implant structure. Under these conditions, the measured input power was $P_i = \SI{6.4}{\watt}$, and the voltage regulator was properly biased, resulting in an output power of $P_o = \SI{3.3}{\milli\watt}$. Consequently, the power transfer efficiency was $\eta = 0.05\%$.

\section{Conclusions}
This article presents an analysis of a wireless power transfer system based on inductive coupling to an electronic device located inside a cavity of a Ti alloy implant or a similar conductive material, primarily employing solenoid-type coils. In typical applications, there are inherent constraints related to geometry and the use of rigid, non-flexible materials, which often require dedicated engineering solutions. A theoretical analysis is presented to illustrate the effects of induced currents in the conductive parts on the receiver circuit, as well as their dependence on frequency and the challenges imposed on power transmission itself and on circuit tuning.

Several representative geometries are simulated to illustrate the effect of eddy currents in the conductive prosthesis on the receiver circuit, and the magnetic coupling coefficient is calculated. To mitigate the effects of such currents, several techniques are proposed: (a) the inclusion of discontinuities in the conductor through drilled holes, aiming to increase the electrical resistance; (b) the use of a high-permeability magnetic film to concentrate the magnetic flux and reduce the magnetic field density at the conductive surface, thereby decreasing the magnitude of the induced currents; (c) the complete interruption of conductive paths that form current loops in the conductive material; and (d) a change in geometry toward planar spiral coils.

An experimental setup was assembled with a load drawing 3.3~V and 1~mA DC. Each of the proposed techniques was tested and monitored. It was observed that the mitigation techniques based on the inclusion of discontinuities in the conductive implant and the use of a high-permeability magnetic film, although reducing the intensity of eddy currents and increasing the magnetic coupling coefficient $k$, have a practically negligible overall effect. Although the use of planar coils depends on application-specific constraints as well as on the overall dimensions involved,  the magnetic coupling coefficient $k$ increased from 0.005 to $k=0.01$ the experimental case considered. Finally, considering the solenoid geometry with a high permeability core with complete interruption of the conductive loop formed by the cavity in Ti revealed to be the best-case, reaching $k=0.02$. This resulted in an efficiency variation, in the worst case, from 0.002\% to 2.2\%. Although this efficiency may still appear low, in the first case a transmitter coil current of 2.8~A was required, while in the best case only 600~mA was necessary.

\section*{Acknowledgments}

This study was financed in part by the Coordena\c{c}\~{a}o de Aperfei\c{c}oamento de Pessoal de N\'ivel Superior - Brasil (CAPES) Finance Code 001, by INCT-NAMITEC (CNPq n. 406193/2022- 3) and by FAPERGS (Funda\c{c}\~{a}o de Amparo a Pesquisa do Estado do RS), processo 23/2551-0002201-0.





%

\bibliographystyle{IEEEtran}

\bibliography{implante} 	

%
%

\end{document}